\documentclass[11pt,a4paper,sans,english]{article}
\usepackage{amsmath}
\usepackage[titletoc]{appendix}
\usepackage{amssymb}
\usepackage{mathtools}
\usepackage{subcaption}
\usepackage{graphicx, caption, subcaption}
\usepackage{url}
\usepackage{float}
\usepackage{textcomp}
\usepackage[perpage]{footmisc}
\usepackage{tocbibind}
\usepackage{fancyhdr}
\usepackage{pdfpages}

\usepackage{csquotes}

\usepackage{wasysym}
\usepackage[labelfont=bf]{caption}

\usepackage{amsmath}

\usepackage[none]{hyphenat}

\makeatletter
\newcommand{\mathleft}{\@fleqntrue\@mathmargin0pt}
\newcommand{\mathcenter}{\@fleqnfalse}
\makeatother


\usepackage{array}
\newcolumntype{P}[1]{>{\centering\arraybackslash}p{#1}}

\usepackage{dblfloatfix}
\usepackage{amsmath}
\usepackage{amssymb}
\usepackage{mathtools}
\usepackage{graphicx}
\usepackage{url}
\usepackage{listings}
\usepackage{textcomp}
\usepackage[perpage]{footmisc}
\usepackage{fancyhdr}

\usepackage{wasysym}
\usepackage[labelfont=bf]{caption}
\usepackage{hyperref}
\hypersetup{colorlinks,
	citecolor=magenta,
	filecolor=red,
	linkcolor=blue,
	linktoc=page,
	urlcolor=green}
\date{}
\title{Monte Carlo Event Generator for the Production and Decay of String Resonances in Proton-Proton Collisions \\ STRINGS Version 1.00}
\author{
  \textbf{Pourya Vakilipourtakalou}\\
  \texttt{University of Alberta}\\
  \\
  \textbf{Douglas M. Gingrich}\\
  \texttt{University of Alberta, TRIUMF}\\
}
\begin{document}
\maketitle

\begin{abstract}
\noindent We describe a Monte Carlo event generator for the production and decay of first and second string resonances through 2 $\rightarrow$ 2 partonic and also 2-parton $\rightarrow$ $\gamma$-parton scatterings in proton-proton collisions - STRINGS version 1.00. This generator is also capable of producing QCD diparton processes. STRINGS is written in Python 2 and can be interfaced to common hadronization programs using the Les Houches Accord.
\end{abstract}

\pagebreak
\tableofcontents
\pagebreak

\section{Event Generator} \label{generator}

Using an intersecting D-branes model \cite{dbrane1, dbrane2} in large extra dimensions \cite{nima1, nima2} could lead to low-scale string theories \cite{lowscale}, in which the string scale $M_s$ is of the order of a few TeV. In the D-brane formulation of low-scale string theory, string resonances (Regge excitations) can be produced in proton-proton collisions through 2-parton scatterings \cite{venez, regge}, and furthermore, in the limit $M_s \rightarrow \infty$ these scattering amplitudes match the ones derived in QCD.\\

We present a Monte Carlo event generator for the production and decay of the first and second string resonances, and also for QCD tree-level scatterings, in proton-proton collisions, such that colour, quark flavour and electric charge are conserved. We consider the production of the  first and second resonances, and QCD diparton processes through  2~$\rightarrow$ 2 partonic scatterings, while 2-parton $\rightarrow$ $\gamma$-parton scatterings can only produce first string resonance. Interference terms between the resonances are not considered in this generator. The subprocesses, that are used in this generator, are as follows.\\

\noindent 2 $\rightarrow$ 2 partonic scattering:\\

\begin{equation}
gg \rightarrow gg,
\label{gggg}
\end{equation}

\begin{equation}
gg \rightarrow q \bar{q},
\end{equation}

\begin{equation}
gq \rightarrow gq,
\end{equation}

\begin{equation}
g \bar{q} \rightarrow g \bar{q},
\end{equation}

\begin{equation}
q \bar{q} \rightarrow gg,
\label{qqbar}
\end{equation}






\noindent 2-parton $\rightarrow$ $\gamma$-parton scattering:\\

\begin{equation}
gg \rightarrow g \gamma,
\end{equation}

\begin{equation}
 gq \rightarrow q \gamma.
\label{gqgamma}
\end{equation}
\\



The event generator works based on the cross-sections and decay widths described in Ref. \cite{crosswid}. The cross-section of the proton-proton collision is given by the convolution of the parton distribution functions (PDFs) with the spin averaged squared partonic scattering amplitudes $|M(ij \rightarrow kl)|^2$. By defining $y$ and $Y$ in terms of the rapidities of the outgoing partons $y_1$ and $y_2$ as\\

\begin{equation}
\begin{split}
Y &\equiv \dfrac{1}{2}(y_1 + y_2),\\
\\
y &\equiv \dfrac{1}{2}(y_1 - y_2),
\end{split}
\end{equation}

\noindent the three-dimensional differential cross-section in terms of $y$, $Y$ and the invariant mass of the partons $M$ can be written as \cite{proofbest}\\

\begin{equation}
\dfrac{d^3\sigma}{dMdydY} = \sum_{ij} \dfrac{f_i(x_a, Q)f_j(x_b, Q) |M(ij \rightarrow kl)|^2}{16 \pi M s},
\label{eq1}
\end{equation}
\\

\noindent where, $s$ is the centre of mass energy of the proton-proton collision. The summation is over the flavours of the incoming partons $(i,j)$ and $f_{i,j}$ are the PDFs for the incoming partons evaluated at the energy scale of the interaction $Q$. The fractions of the incoming protons' four-momenta, that are carried by the incoming partons, are shown by $x_a$ and $x_b$, which can be written as functions of $M$ and $Y$ \cite{proofbest}

\begin{equation}
\begin{split}
x_a &= \sqrt{\tau}e^Y, \\
\\
x_b &= \sqrt{\tau}e^{-Y},
\end{split}
\label{xs}
\end{equation}
\\

\noindent where,\\

\begin{equation}
\tau = \dfrac{M^2}{s}.
\end{equation}
\\

The summation in Eq. (\ref{eq1}) is performed over all incoming parton flavours. Indeed, for an individual subprocess, e.g. $ij \rightarrow kl$, the  general three-dimensional differential cross-section in Eq. (\ref{eq1}) is written as\\

\begin{equation}
F_{ij}(M,Y,y) \equiv \dfrac{f_i(x_a, Q)f_j(x_b, Q) |M(ij \rightarrow kl)|^2}{16 \pi M s},
\label{eq2}
\end{equation}
\\

\noindent which means that the total three-dimensional cross-section is written as\\

\begin{equation}
\dfrac{d^3\sigma}{dMdydY} = \sum_{ij} F_{ij}(M,Y,y).
\label{eq222}
\end{equation}
\\

Since $x_a$ and $x_b$ are functions of $Y$, Eq. (\ref{xs}), the distribution function $F_{ij}(M, Y, y)$ in Eq. (\ref{eq222}) can be used to generate events, and each generated event is specified by three independent variables ($M, Y, y$). The generator calculates all of the kinematic variables of the partons from $(M, Y, y)$, as will be discussed later, and saves them in a LHE (Les Houches Event) file \cite{lhefile}. The first step in the generator is to perform an integration of $F_{ij}(M, Y, y)$ over $y$ and $Y$ to get the one-dimensional differential cross-section as a function of the invariant mass for the individual subprocesses as follows \cite{proofbest}\\

\begin{equation}
G_{ij}(M) \equiv \dfrac{d\sigma}{dM} = \int_{-Y_{\mathrm{max}}}^{Y_{\mathrm{max}}}dY\int_{-(y_{\mathrm{max}}-|Y|)}^{y_{\mathrm{max}}-|Y|} F_{ij}(M,Y,y)dy, 
\label{eq3}
\end{equation}
\\

\noindent in which, $y_{\mathrm{max}}$ is the upper limit on the absolute value of the rapidities of the outgoing partons $|y_1|, |y_2| < y_{\mathrm{max}}$ and $Y_{\mathrm{max}}$ is a function of $y_{\mathrm{max}}$ \cite{proofbest}\\

\begin{equation}
Y_{\mathrm{max}} = \mathrm{min}\{\mathrm{ln}(1/\sqrt{\tau}, y_{\mathrm{max}})\}.
\end{equation}
\\

After performing the integral in Eq. (\ref{eq3}), $G_{ij}(M)$ can be regarded as a distribution function to generate the invariant mass of the event and also the type of subprocess. For example, for the 2 $\rightarrow$ 2 partonic  scattering, Eqs. (\ref{gggg}-\ref{qqbar}), considering the flavours of the incoming partons, there are 20 possible subprocesses. A random invariant mass is generated in the specified interval and for that invariant mass, there are 20 values for the differential cross-sections. A second uniform random number is generated between the maximum and minimum values of the set of the differential cross-sections and depending on where the second random number lands, the subprocess is determined. Determining the subprocess means that the flavours of the incoming partons are determined as well. A third uniform random number is generated between the minimum and maximum of $G_{ij}(M)$ in the specified invariant mass interval and if the third random number is smaller than $G_{ij}$ evaluated at the random $M$, the random invariant mass $M$ is kept. Otherwise, we restart the algorithm from the beginning.\\

Once the type of the scattering subprocess and the invariant mass of the partons are specified, they can be inserted into $F_{ij}(M, Y, y)$ on the right-hand side of the Eq. (\ref{eq3}), to generate $y$ and $Y$. This procedure is the same as generating the invariant mass using $G_{ij}(M)$ except it is a two-dimensional distribution function, which depends on $Y$ and $y$. Knowing that the maximum and minimum values of $F_{ij}(M, Y, y)$, for a given $M$, are $F$($M$, 0, 0) and 0, respectively, makes it faster to generate $y$ and $Y$. Uniform random $y$ and $Y$ are generated in the intervals determined by $y_{\mathrm{max}}$, given by Eq. (\ref{eq3}). A third uniform random number is generated between the minimum and maximum values of $F_{ij}$ and if the third random number is smaller than $F_{ij}$ evaluated at the random $Y$ and $y$, they are kept. Otherwise, we restart the algorithm from the beginning, i.e. the generation of the random invariant mass and determination of the subprocess.\\

Having $M$, $Y$ and $y$, together with the assumption that the incoming protons collide along the $z$-axis with the following four-momenta\\

\begin{equation}
\begin{split}
P_a &= (s/2, 0, 0, s/2),\\
\\
P_b &= (s/2, 0, 0, -s/2),
\end{split}
\end{equation}
\\

\noindent the kinematic variables of the outgoing partons can be calculated, as listed below.\\

\noindent The rapidities of the outgoing partons are\\

\begin{equation}
\begin{split}
y_1 &= Y + y,\\
\\
y_2 &= Y - y.
\end{split}
\end{equation}

\noindent The magnitude of the transverse momentum of the outgoing partons are\\

\begin{equation}
p_{T} = \dfrac{M}{2\mathrm{cosh} y}.
\end{equation}
\\

\noindent The $x$- and $y$-components of the outgoing partons' three-momentum are\\

\begin{equation}
\begin{split}
p_{x_1} &= p_{T}\mathrm{cos} \phi,\\
\\
p_{y_1} &= p_{T}\mathrm{sin} \phi,\\
\\
p_{x_2} &= -p_{x_1},\\
\\
p_{y_2} &= -p_{y_1},
\end{split}
\end{equation}
\\

\noindent where, due to the symmetry of the collision, the azimuthal angle $\phi$ is a uniform random number between $0$ and $2\pi$.\\

\noindent The energies of the outgoing partons are

\begin{equation}
\begin{split}
E_1 &= p_{T}\mathrm{cosh} y_1,\\
\\
E_2 &= p_{T}\mathrm{cosh} y_2.
\end{split}
\end{equation}
\\

\noindent The $z$-components of the outgoing partons' three-momentum are

\begin{equation}
\begin{split}
p_{z_1} &= E_1\mathrm{tanh} y_1,\\
\\
p_{z_2} &= E_2\mathrm{tanh} y_2.
\end{split}
\end{equation}
\\

The QCD running coupling constant, which appears in all of the scattering amplitudes, is given by \cite{PDG}\\

\begin{equation}
\dfrac{1}{\alpha(Q)} = \dfrac{1}{\alpha(M_Z)}+\dfrac{7}{2\pi}\text{ln}\dfrac{Q}{M_Z},
\label{running}
\end{equation}
\\

\noindent where, $Q$ is the scale at which the coupling constant is calculated and $M_Z$ is the mass of the Z boson. \\

The flavours of the incoming partons are determined at the stage where we determine the type of the subprocess and, since flavour is conserved in our model, one can determine the flavours of the outgoing partons except in the $gg \rightarrow q \bar{q}$ subprocess. As soon as this subprocess is selected, the outgoing partons can have any of the six flavours with the same probability.\\

\begin{figure}[h] 

	\begin{subfigure}{0.48\textwidth}
		\includegraphics[width=\linewidth]{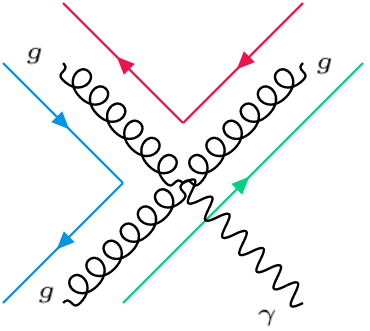}
		\caption{$gg \rightarrow g\gamma$}
	\end{subfigure}\hspace*{\fill}
	\begin{subfigure}{0.48\textwidth}
		\includegraphics[width=\linewidth]{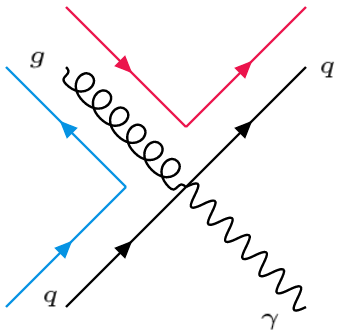}
		\caption{$gq \rightarrow q\gamma$}
	\end{subfigure}
	
	\caption{Colour flow diagrams for different subprocesses of the 2-parton~$\rightarrow$~$\gamma$-parton scattering.}
	\label{fig1}
\end{figure}

\begin{figure}[h] 

	\begin{subfigure}{0.48\textwidth}
		\includegraphics[width=\linewidth]{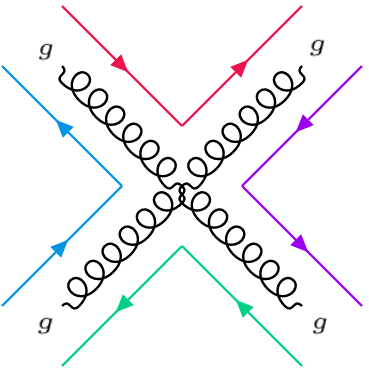}
		\caption{$gg \rightarrow gg$}
	\end{subfigure}\hspace*{\fill}
	\begin{subfigure}{0.48\textwidth}
		\includegraphics[width=\linewidth]{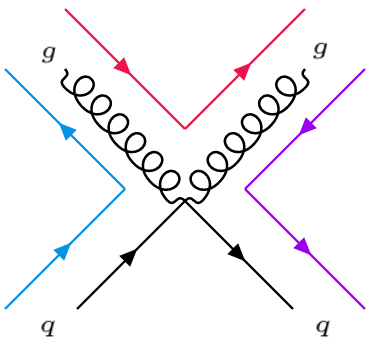}
		\caption{$gq \rightarrow gq$}
	\end{subfigure}
	
	\medskip
	\begin{subfigure}{0.48\textwidth}
		\includegraphics[width=\linewidth]{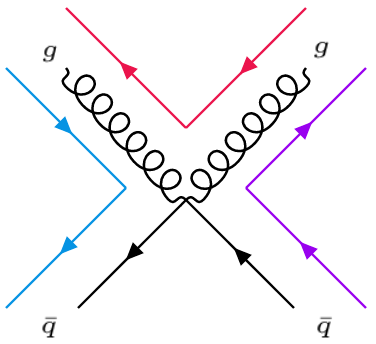}
		\caption{$g \bar{q} \rightarrow g \bar{q}$}
	\end{subfigure}\hspace*{\fill}
	\begin{subfigure}{0.48\textwidth}
		\includegraphics[width=\linewidth]{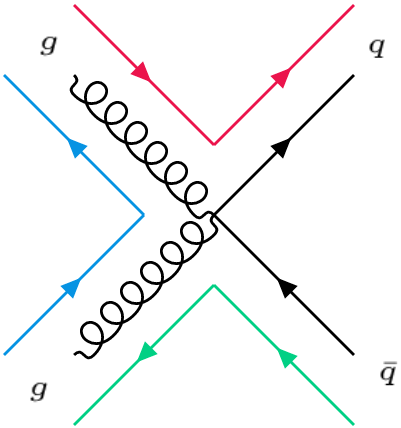}
		\caption{$gg \rightarrow q \bar{q}$}
	\end{subfigure}
	
	\medskip
	\begin{subfigure}{0.48\textwidth}
		\includegraphics[width=\linewidth]{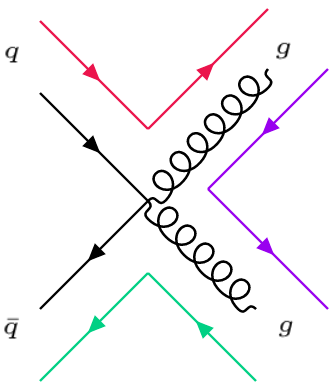}
		\caption{$q \bar{q} \rightarrow gg$}
	\end{subfigure}\hspace*{\fill}
	
	\caption{Colour flow diagrams for different subprocesses of the 2 $\rightarrow$ 2 parton scattering.}
	\label{fig2}
\end{figure}

In the LHE file, one needs to specify the colour flow in each event and in order to take care of the conservation of the colour in the hadronization process, we used four numbers in the LHE file, 101, 102, 103 and 104. As in figures~\ref{fig1} and \ref{fig2}, these numbers are represented by a colour (not to be confused with the QCD colours), and the conservation of colour is satisfied if the summation of colour numbers on the left-hand side of the vertex is equal to that of the right-hand side of the vertex. These numbers are saved in the LHE file.\\
\\
\\

\clearpage

\section{How to Use the Generator}
\label{usage}

The STRINGS generator has several parameters which can be given as inputs to the program. We used the package \texttt{argparse} which is a parser for the command-line options. For each input that is passed to the generator, we used two versions of parameters. The user may run the generator by typing one of the following lines.\\

\noindent \texttt{python STRINGS.py -V1argument argumentValue}\\

\noindent or\\

\noindent \texttt{python STRINGS.py --V2argument argumentValue}\\

\noindent where, \texttt{V1argument} and \texttt{V2argument} are the two versions of the parameters listed below, and \texttt{argumentValue} is the value of that parameter. If any of the parameters are not set by the above command line, they will preserve their default values. In the following, we provide all of the parameters used in the generator together with their default values and units, if applicable. 

\subsection{Inputs}

Users can easily change the generator parameters using the following arguments. The first (second) parameter represents the first (second) version of the input parameter. The default values are shown as (D=...).\\

\noindent \texttt{-RandGenSeed}, \texttt{--RandGenSeedValue} : (D = 123456) The seed for the random number generator. In order to get different sequences of generated events, this seed should be changed by the user for each run.\\

\noindent \texttt{-COME}, \texttt{--COMEvalue} (GeV): (D = 13000) Centre of mass energy of the incoming protons.\\
\clearpage

\noindent \texttt{-Ms}, \texttt{--Msvalue} (GeV): (D = 7000) The string scale of the string theory. The string scale can be chosen to have any positive value.\\

\noindent \texttt{-MinMass}, \texttt{--MinMassvalue} (GeV): (D = 6000) The lower bound for the invariant mass of the partons. This value should always be greater than zero.\\

\noindent \texttt{-MaxMass}, \texttt{--MaxMassvalue} (GeV): (D = 8000) The upper bound for the invariant mass of the partons. This value should always be smaller than the centre of mass energy. Otherwise, the fraction of the carried momentum by one of the partons would be equal or greater than one.\\

\noindent \texttt{-ymax}, \texttt{--ymaxvalue}: (D = 2.5) The upper bound for the rapidities (not pseudo-rapidity) of the outgoing partons. In order to set this value to infinity the user can use $y_{\mathrm{max}}=-1$~.\\

\noindent \texttt{-Number}, \texttt{--Numbervalue}: (D = 10000) Number of events to be generated.\\

\noindent \texttt{-Coupling}, \texttt{--Couplingvalue}: (D = $-1$) The QCD coupling constant \- $\alpha_s=g_s^2/4\pi$ that is used in the scattering amplitudes. The default value for this variable is $-1$, which is just a terminology to specify that the running coupling constant is being used and it is calculated at the QCD scale. The user can change this to a fixed coupling constant by changing the default value to the intended value.  \\

\noindent \texttt{-CouplingScale}, \texttt{--CouplingScalevalue}: (D = \texttt{Ms}) The QCD scale at which the running coupling constant is calculated. This is used only if the running coupling constant is used, i.e. \texttt{Coupling} $ = -1$. \\

\noindent \texttt{-PDFSet}, \texttt{--PDFSetvalue}: (D = $\texttt{\textquotedblright cteq6l1\textquotedblright}$) The parton distribution function (PDF) set. This generator uses the PDF sets of the LHAPDF \cite{lhapdf}. This parameter should be given to the generator as a string type. Some common examples of the PDF sets are \texttt{\textquotedblright CT10\textquotedblright}, \texttt{\textquotedblright CT14lo\textquotedblright} and \texttt{\textquotedblright NNPDF23LO\textquotedblright}; leading order PDFs are recommended.\\

\noindent \texttt{-PDFScale}, \texttt{--PDFScalevalue} (GeV): (D = \texttt{Ms}) The QCD scale at which the PDFs are evaluated. \\

\clearpage

\noindent STRINGS can produce events for the first and second string resonances, through different scattering processes, and also the QCD tree-level diparton process. Since more than one type of event may be generated, three booleans are used to specify the intended type of event generation.\\

\noindent \texttt{-QCDCoeff}, \texttt{--QCDCoeffvalue}: (D = \texttt{false}) A boolean which specifies the event generation for the QCD tree-level diparton production. \texttt{true} (\texttt{false}) means that the QCD tree-level scattering events are (are not) generated. \\

\noindent \texttt{-FirstStringCoeff}, \texttt{--FirstStringCoeffvalue}: (D = \texttt{true}) A boolean which specifies event generation for the first string resonance through 2~$\rightarrow$~2 parton and 2-parton $\rightarrow$ $\gamma$-parton scatterings. \texttt{true} (\texttt{false}) means that the production of the first string resonance through these types of scattering is (is not) in the event generation. \\

\noindent \texttt{-SecondStringCoeff}, \texttt{--SecondStringCoeffvalue}: (D = \texttt{false}) A boolean which specifies event generation for the second string resonance through 2~$\rightarrow$~2 parton scatterings. \texttt{true} (\texttt{false}) means that the production of the second string resonance is (is not) in the event generation. Please note that the second resonance is not implemented for 2-parton $\rightarrow$ $\gamma$-parton scatterings.\\

\noindent In order to produce pure QCD events, for example, the user should set \texttt{QCDCoeff} to \texttt{true} and other types of event generation to \texttt{false}.\\

\noindent There are seven subprocesses that are used in the generator, Eqs. (\ref{gggg}-\ref{gqgamma}). Each of these subprocesses are specified by a boolean such that \texttt{true} (\texttt{false}) as their value means that those subprocesses are (are not) considered in the production of the string resonances or also in the QCD diparton production. These variables and their default values are shown below\\

1: \texttt{-gg2gg}, \texttt{--gg2ggvalue} = \texttt{true}\\

2: \texttt{-gg2qqbar}, \texttt{--gg2qqbarvalue} = \texttt{true}\\

3: \texttt{-gq2gq}, \texttt{--gq2gqvalue} = \texttt{true}\\

4: \texttt{-gqbar2gqbar}, \texttt{--gqbar2gqbarvalue} = \texttt{true}\\

5: \texttt{-qqbar2gg}, \texttt{--qqbar2ggvalue} = \texttt{true}\\

6: \texttt{-gg2gGamma}, \texttt{--gg2gGammavalue} = \texttt{false}\\

7: \texttt{-gq2qGamma}, \texttt{--gq2qGammavalue} = \texttt{false}\\

\noindent In the LHE file, for each event, the subprocess for that event is specified by a number (1 through 7) as given above.\\

\noindent Quark masses can be specified by the user. Their default values are given below\\

\noindent \texttt{-dMass}, \texttt{--dMassvalue} (GeV): (D = 5 $\times$ 10$^{-3}$) Mass of the down quark .\\
\\
\noindent \texttt{-uMass}, \texttt{--uMassvalue} (GeV): (D = 2 $\times$ 10$^{-3}$) Mass of the up quark.\\
\\
\noindent \texttt{-sMass}, \texttt{--sMassvalue} (GeV): (D = 1 $\times$ 10$^{-3}$) Mass of the strange quark.\\
\\
\noindent \texttt{-cMass}, \texttt{--cMassvalue} (GeV): (D = 1.27) Mass of the charm quark.\\
\\
\noindent \texttt{-bMass}, \texttt{--bMassvalue} (GeV): (D = 4.4) Mass of the bottom quark.\\
\\
\noindent \texttt{-tMass}, \texttt{--tMassvalue} (GeV): (D = 1.72 $\times$ 10$^{2}$) Mass of the top quark.\\
\\

Here we provide an example on how to run the generator and change a parameter. For example, if the user intends to generate events for the production of the first and second string resonances, since \texttt{FirstStringCoeff} is set to be \texttt{true} by default, they only need to set \texttt{SecondStringCoeff} to \texttt{true} by typing either of the following lines.\\

\noindent \texttt{python STRINGS.py -SecondStringCoeff true}\\

\noindent or\\

\noindent \texttt{python STRINGS.py ---SecondStringCoeffvalue true}\\

\noindent Sometimes, the user needs to change many parameters and it would not be convenient to pass them through command line. There is a bash script \texttt{run.sh} that is included in the STRINGS package, which includes all of the input parameters, initially set to their default values. Users can change the parameters to have the intended values and type the following command line.\\

\noindent \texttt{source run.sh}\\

\clearpage

\subsection{Code Flow}

The generator uses the built-in packages of Python such as \texttt{math}, \texttt{random}, and \texttt{argparse}, together with the installed \texttt{lhapdf}. At the beginning of the main code the following lines appear. \\

\noindent \texttt{import lhapdf}\\
\noindent \texttt{import math}\\
\noindent \texttt{import random}\\
\noindent \texttt{import argparse}\\

\noindent \texttt{lhapdf} is used to specify the PDF set, \texttt{math} is used for the basic math operations, \texttt{random} is used for the random generation of numbers; the seed for the random number generator can be set by the user, see Section~\ref{usage}. \texttt{argeparse} is a package which simplifies the way the inputs are passed to the generator. In order to use this package, an object \texttt{parser} is created by the following line\\

\noindent \texttt{parser = argparse.ArgumentParser()}\\

\noindent and the parameters, for example the string scale, are passed to the main function by the following command line.\\

\noindent \texttt{parser.add\_argument(\textquotedblright -Ms\textquotedblright, \textquotedblright --Msvalue\textquotedblright, default=5000)}\\

After specifying the packages, the \texttt{main()} function is defined and takes all of the parameters, which are stated in the \texttt{run.sh} script, as inputs. Basically, everything is defined in this function and after its definition is finished, the function is called with the specified parameters.\\

\noindent The PDF set is defined as follows\\

\noindent \texttt{DistFunc = lhapdf.mkPDF(\textquotedblright PDFSet\textquotedblright, 0)}\\

\noindent in which, \texttt{PDFSet} is given as a string type by the user. This PDF set is called while calculating the differential cross-section as follows\\

\noindent \texttt{DistFunc.xfxQ(partonID, x, Scale)}\\

\noindent in which, \texttt{partonID} is the particles' ID number \cite{PDG}, \texttt{x} is the fraction of the proton's four-momentum carried by the parton and \texttt{Scale} is the QCD scale at which the PDFs are evaluated, i.e. \texttt{PDFScale}.\\
\clearpage

There are seven functions inside the main function corresponding to the subprocesses. These functions are shown below.\\

\noindent \texttt{def MonteCarlogggg(parameters, *args)}\\

\noindent \texttt{def MonteCarloggqqbar(parameters, *args)}\\

\noindent \texttt{def MonteCarlogqgq(parameters, *args)}\\

\noindent \texttt{def MonteCarlogqbargqbar(parameters, *args)}\\

\noindent \texttt{def MonteCarloqqbargg(parameters, *args)}\\

\noindent \texttt{def MonteCarlogggGamma(parameters, *args)}\\

\noindent \texttt{def MonteCarloqgqGamma(parameters, *args)}\\

These functions are used to calculate the differential partonic cross-sections for the corresponding subprocesses and also to generate $Y$ and $y$. The argument \texttt{parameters} in these functions is an array which consists of $Y$ and $y$, and \texttt{args} is a tuple which consists of the invariant mass and the IDs of the incoming partons. They are used in these function as follows.\\

\noindent \texttt{Y, y = parameters}\\

\noindent \texttt{M, FstpartonID, ScndpartonID = args}\\

\noindent In each of these functions, the spin averaged matrix element squared for each process is calculated and they are added up to make the total scattering amplitude.\\

\noindent \texttt{MSquareTotal} = \texttt{QCDCoeffint*MSquareQCD + }\\
\hspace*{2.85 cm} \texttt{FirstStringCoeffint*MSquareFirstString +}\\
\hspace*{2.85 cm} \texttt{SecondStringCoeffint*MSquareSecondString}\\

\noindent \texttt{QCDCoeffint}, \texttt{FirstStringCoeffint} and \texttt{SecondStringCoeffint} are the integers corresponding to the three boolean parameters specifying the type of the event generation, i.e. \texttt{QCDCoeff}, \texttt{FirstStringCoeff} and \texttt{SecondStringCoeff}, respectively; if the user sets the booleans to be \texttt{true} (\texttt{false}), the corresponding integer variable will be $1$ ($0$). If the user sets one of these booleans as to be \texttt{false}, the contribution from the corresponding scattering amplitude is zero. Eventually, depending on which subprocesses are turned on by the user, the corresponding functions are used to calculate the right-hand side of Eq. (\ref{eq3}) in the event generation loop. The event generation loop, a \texttt{while} loop, is set to make sure the intended number of events are generated. Before the event generation loop, a LHE file is created to save the events;\\

\noindent \texttt{fh = open("events.lhe", "w")}\\

\noindent and after the event generation is done, the cross-section is calculated and put in the LHE file. All of the input variables are shown on the screen and also saved in the LHE file.\\





			
	
		




	
		



\section{Performance}

\subsection{Accuracy}

To each event, that is generated in the event generation loop, a cross-section is assigned as follows\\

\begin{equation}
\sigma_i = F_i \times (M_{\mathrm{max}} - M_{\mathrm{min}}),
\end{equation}
\\

\noindent where, $F_i$ is the differential cross-section for the event and $M_{\mathrm{max}}$ ($M_{\mathrm{min}}$) is the upper (lower) bound for the invariant mass. The total cross-section is calculated as the mean value of the individual cross-sections as\\

\begin{equation}
\sigma_{\mathrm{total}} = \sum_{i=1}^{i=N}\dfrac{\sigma_i}{N},
\end{equation}
\\

\noindent where, $N$ is the total number of generated events. The uncertainty in the total cross-section can be stated as\\

\begin{equation}
E = \dfrac{S}{\sqrt{N}},
\end{equation}
\\

\noindent where, $S$ is the standard deviation defined by \\

\begin{equation}
S = \sqrt{\sum_{i=1}^{N}\dfrac{\sigma_i^2}{N} - \left(\sum_{i=1}^{N}\dfrac{\sigma_i}{N}\right)^2}.
\end{equation}
\\

\begin{figure}[H]
  \centering
  \includegraphics[width=\linewidth]{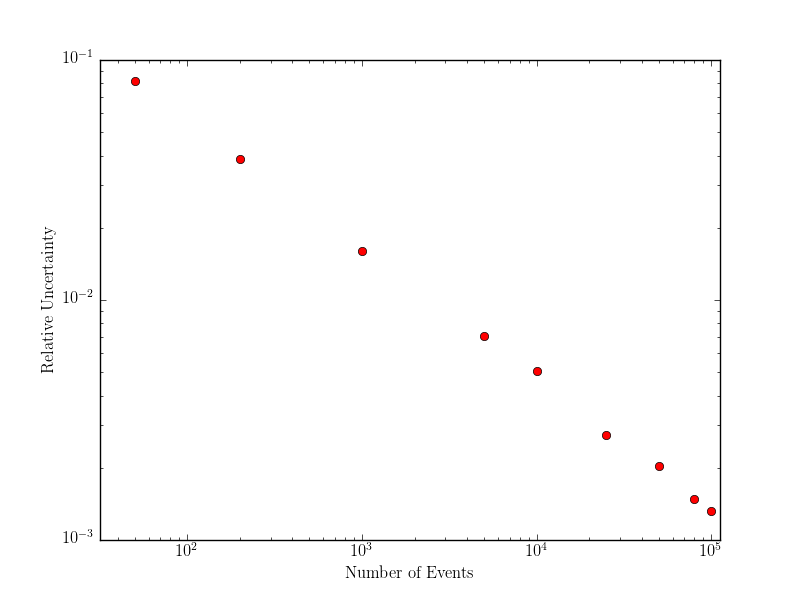}
  \caption{Estimated relative uncertainty in the total cross-section versus number of generated events when using the default parameters.}
  \label{fig:error}
\end{figure}

We ran the generator for different number of events using the default parameters. Figure \ref{fig:error} shows the estimated relative uncertainty in the cross-section versus number of events generated.

\subsection{Invariant Mass Interval}

Due to the convolution of the scattering amplitudes with the PDFs in \- Eq.~(\ref{eq3}), the differential cross-sections for different subprocesses drop rapidly as the invariant mass increases, and events with higher invariant masses are less likely to be generated. Thus, the larger the invariant mass interval is, the more events are rejected in the event generation loop, that causes a decrease in the performance of the generator. Knowing that the scattering amplitudes for the production of the string resonances are of the form of the Breit-Wigner distribution \cite{breit-wigner}, we suggest that the invariant mass interval $\Delta M$ around the resonance $M_s$ be no more than four times the decay width of the resonance, which covers more than 99\% of the area under the resonance. Table (\ref{table1}) contains the suggested invariant mass intervals for different string scales, while generating events for the first string resonance with the default parameters. If the user wants to generate events for larger invariant mass intervals, it would be better to break the intended invariant mass interval into smaller slices and run the generator for the smaller intervals. The slices can then be combined to build up the entire distribution by weighting each by its production cross-section.\\

\begin{table}[h]
\centering
		\begin{tabular}{ | P{3cm} | P{3cm} | }
			\hline
			 $M_s$ (GeV) & $\Delta M$ (GeV)\\
			\hline
			1000      & $\pm$ 150 \\
			3000      & $\pm$ 300 \\
			5000      & $\pm$ 450  \\
	        7000      & $\pm$ 600  \\
		    9000      & $\pm$ 750 \\
		    11000     & $\pm$ 900 \\
			\hline
 \end{tabular}	
 \caption{Suggested invariant mass intervals for different string scales, which cover more than 99\% of the area under the first string resonance when using the default parameters.}
\label{table1}
\end{table}

\subsection{Timing}

\begin{figure}[H]
  \centering
  \includegraphics[width=\linewidth]{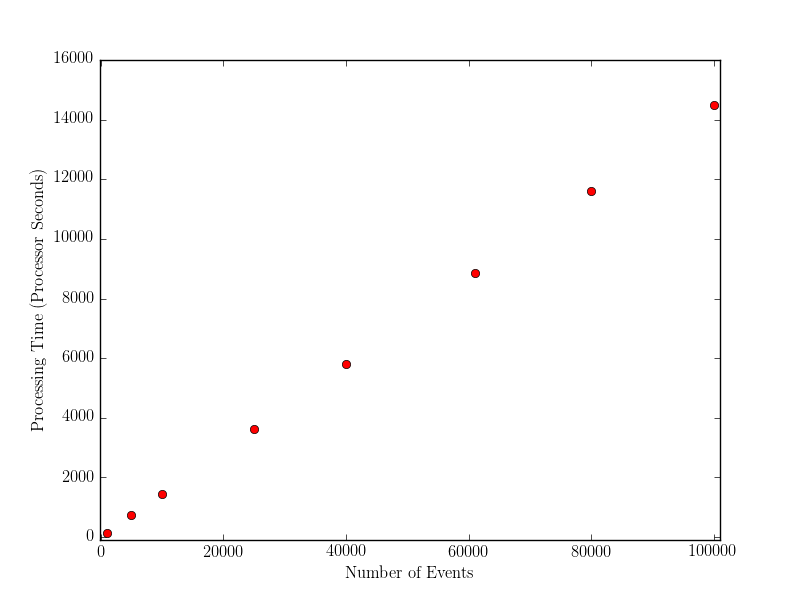}
  \caption{The processing time versus number of generated events for the production of the first string resonance at $M_s = 7000$ GeV, while all of the subprocesses are on.}
  \label{fig:1}
\end{figure}

We ran the generator for the invariant mass interval $M$ = [$6000$ - $8000$] GeV with all of the subprocesses, i.e. 2 $\rightarrow$ 2 parton and 2-parton $\rightarrow$ $\gamma$-parton scatterings, contributing in the production of the first string resonance (QCD is not included) at string scale \- $M_s$~=~$7000$~GeV and measured the time in processor seconds for the three stages of initialization, event generation and termination. The processor that we used was an \texttt{Intel E5-2609 v4} with \texttt{64 MB} of memory. Initialization and termination of the generator took 0.010 and 0.059 processor seconds, respectively. Figure \ref{fig:1}, shows the time it took for the processor to perform the event generation loop versus the number of generated events. We also fitted a line to the data and the equation is given as

\begin{equation}
T = 0.145 \times N + 0.069\ [\mathrm{Processor}\ \mathrm{Seconds}],
\end{equation}

\noindent where, $T$ is the processing time in processor seconds and $N$ is the number of the generated events.

\section{Installation and Availability}

The STRINGS generator is written in Python 2, which is available on most Linux systems. STRINGS uses the PDF sets of the LHAPDF library \cite{lhapdf} and it should be installed with Python enabled. For the instructions on how to install LHAPDF visit \texttt{https://lhapdf.hepforge.org}; all of these instructions are included in a README file, together with an example shell script \texttt{setup.sh} to configure the environment variables, as part of the STRINGS package. \texttt{setup.sh} should be sourced once before runing the STRINGS. The source code for the STRINGS generator can be obtained from the HepForge site: \href{https://strings.hepforge.org/}{https://strings.hepforge.org/}.


\clearpage

\begin{thebibliography}{9}


\bibitem{dbrane1}
M. R. Garousi and R. C. Myers, Superstring Scattering from D-Branes, Nucl. Phys. B 475 (1996) 193, [\href{https://arxiv.org/abs/hep-th/9603194}{ArXiv:hep-th/9603194}].

\bibitem{dbrane2}
A. Hashimoto and I. R. Klebanov, Decay of Excited D-branes, Phys. Lett. B 381 (1996) 437, [\href{https://arxiv.org/abs/hep-th/9604065}{ArXiv:hep-th/9604065}].

\bibitem{nima1}
N. Arkani-Hamed, S. Dimopoulos and G.R. Dvali, The Hierarchy Problem and New Dimensions at a Millimeter, Phys. Lett. B 429 (1998) 263, [\href{https://arxiv.org/abs/hep-ph/9803315}{ArXiv:hep-ph/9803315}].

\bibitem{nima2}
I. Antoniadis, N. Arkani-Hamed, S. Dimopoulos and G. R. Dvali, New Dimensions at a Millimeter to a Fermi and Superstrings at a TeV, Phys. Lett. B 436 (1998) 257, [\href{https://arxiv.org/abs/hep-ph/9804398}{ArXiv:hep-ph/9804398}].

\bibitem{lowscale}
I. Antoniadis, A Possible new dimension at a few TeV, Phys. Lett. B 246 (1990) 377.

\bibitem{venez}
G. Veneziano, Construction of a Crossing-Symmetric, Regge-behaved Amplitude for Linearly Rising Trajectories, Nuovo Cim. A 57 (1968) 190.

\bibitem{regge}
R. L. Thews, Regge Poles in Resonance Production, Phys. Rev. 155 (1967) 1624.

\bibitem{crosswid}
S. Cullen, M. Perelstein and M. E. Peskin, TeV Strings and Collider Probes of Large Extra Dimensions, Phys. Rev. D 62 (2000) 055012, [\href{https://arxiv.org/abs/hep-ph/0001166}{ArXiv:hep-ph/0001166}].


\bibitem{proofbest}
E. Eichten, I. Hinchliffe, K. D. Lane and C. Quigg, Supercollider Physics, Rev. Mod. Phys. 56 (1984) 579.

\bibitem{65}
 D. Lust, O. Schlotterer, S. Stieberger and T. Taylor, The LHC String Hunter’s Companion (II): Five-Particle Amplitudes and Universal Properties, Nucl. Phys. B 828 (2010) 139, [\href{https://arxiv.org/abs/0908.0409}{ArXiv:hep-th/0908.0409}].
 
\bibitem{PDG}
C. Patrignani et al., Particle Data Group, Chin. Phys. C 40 (2016) 100001 and 2017 update, [\href{http://pdg.lbl.gov/}{pdg.lbl.gov}].
 
 
\bibitem{lhefile}
 J. Alwall et al., A Standard Format for Les Houches Event Files, Comput. Phys. Commun. 176 (2007) 300, [\href{https://arxiv.org/abs/hep-ph/0609017}{ArXiv:hep-ph/0609017}].
 
 \bibitem{breit-wigner}
L. A. Anchordoqui, H. Goldberg and T.R. Taylor, Phys. Lett. B 668 (2008) 373, [\href{https://arxiv.org/abs/0806.3420}{ArXiv:hep-ph/0806.3420}].
 
\bibitem{lhapdf}
A. Buckley et al., LHAPDF6: Parton Density Access in the LHC Precision Era, Eur. Phys. J. C 75 (2015) 132, [\href{https://arxiv.org/abs/1412.7420}{ArXiv:hep-ph/1412.7420}].



%









 












\end{thebibliography}
\end{document}